\documentclass[tradiabstract]{aa}
\usepackage{graphicx}
\usepackage[varg]{txfonts}
\usepackage{url}
\usepackage{multirow}
\usepackage{natbib}
\usepackage{color}
\usepackage{longtable}
\usepackage{pdflscape} 

\bibpunct{(}{)}{;}{a}{}{,} 

\def \FLL {\emph{Fermi}-LAT~}

\begin{document}

\title{A MST catalogue of $\gamma$-ray source candidates  above 10 GeV\\
and at Galactic latitudes higher than 20\degr}
\author{R. Campana
		\inst{1},
	E. Massaro
              \inst{2,3},
	E. Bernieri
		\inst{4,5}
}
\institute{INAF/OAS-Bologna, via Piero Gobetti 101, I-40129 Bologna, Italy
\and INAF/IAPS, via Fosso del Cavaliere 100, I-00133 Roma, Italy
\and In Unam Sapientiam, Piazzale A. Moro 2, I-00185 Roma, Italy
\and INFN–Sezione di Roma Tre, via della Vasca Navale 84, I-00146 Roma, Italy
\and Dipartimento di Matematica e Fisica, Universit\`a Roma Tre, Roma, Italy
}           

\date{Received ..., accepted ...}
\titlerunning{9 year MST catalogue of $\gamma$-ray source candidates}
\authorrunning{R. Campana et al.}

%
\abstract{
We describe a catalogue of $\gamma$-ray source candidates, selected using the 
Minimum Spanning Tree (MST) algorithm on the 9-years Fermi-LAT sky (Pass 8) at 
energies higher than 10~GeV. 
The extragalactic sky at absolute Galactic latitudes above 20\degr\ has been 
investigated using rather restrictive selection criteria, resulting in a 
total sample of 1342 sources. 
Of these, 249 are new detections, not previously associated with $\gamma$-ray 
catalogues. A large fraction of them have interesting counterparts, most likely blazars.
In this paper the main results on the catalogue selection and search of 
counterparts are reported.
}
%


\maketitle

\section{Introduction}

The Large Area Telescope \citep[LAT,][]{ackermann12}
onboard the \emph{Fermi} mission is continuously observing 
the sky at energies higher than $\sim$100~MeV since the beginning of August 2008.
It has collected a very large number of photons and enriched our 
knowledge on galactic and extragalactic high-energy astrophysics.
One of the most relevant achievements of this mission is the discovery
of thousands of new celestial sources of $\gamma$-rays, 
reported in several catalogues covering different time
windows, energy ranges and source types. 

Various methods for detecting point-like sources against the photon and instrumental
background are based on the search of local concentrations of $\gamma$-rays.
Besides the traditional Maximum Likelihood (ML) algorithm \citep{mattox96},
other methods include the use of Wavelet Transform \citep{damiani97,ciprini07},
the density-based clustering algorithm DBSCAN \citep{tramacere13,armstrong15}
the Bayesian inferential D$^3$PO algorithm \citep{selig15},
and FermiFAST, recently proposed by \cite{asvathaman17}.

A method, initially proposed by \cite{digesu83} for the analysis of data from the 
COS B satellite \citep{bignami75}, is based on an application of 
the topometric \emph{Minimum} (or \emph{Minimal}) \emph{Spanning Tree}
(hereafter, MST) algorithm, that has its roots in graph theory.
\cite{campana08,campana13} developed this method and introduced new estimators
for the cluster significance to be applied to $\gamma$-ray 2-dimensional images 
where the points correspond to the arrival directions of photons.
The MST method was applied for searching source seeds in preparation of the 
early Fermi catalogues (1FGL, \citealt{1FGL}; 2FGL, \citealt{2FGL}; 1FHL, \citealt{1FHL}).
In a subsequent series of papers 
\citep[][hereafter Papers I--IV]{paperI,paperII,paperIII,paperIV} 
we analyzed the \FLL dataset by means of MST and found several significant $\gamma$-ray 
photon overdensities associated with blazars previously not known as 
high-energy sources.

In this paper we report a new catalogue of $\gamma$-ray clusters, which are robust point-like
source candidates, selected using the MST algorithm in the sky at Galactic latitudes
$|b| > 20\degr$ and at energies higher than 10 GeV.
In this work we considered the Pass~8 \FLL photons observed since the beginning of the mission
up to August 2017, thus covering a period of nine years.
The aim of our search was to enrich the population of weak high-energy sources and to
search for their possible counterparts, particularly among confirmed and candidate blazars, which
consitute the most numerous class of extragalactic $\gamma$-ray sources.

Our sample contains 1342 entries: about 81\% of them have a close spatial association with
known $\gamma$-ray sources reported in public catalogues or surveys, and a further $\sim$8\% can be
associated with known and candidate blazars not previously known as $\gamma$-ray emitters. 

In Section \ref{s:mst_description} the MST method is briefly outlined, while in 
Section \ref{s:data_analysis} the data reduction, MST analysis and the selection 
tools are described. The catalogue content and characteristics are described 
in the following Sections, and in Section \ref{s:conclusions} we discuss our results.

\section{Photon cluster detection with the MST algorithm}\label{s:mst_description}

The Minimum Spanning Tree \citep{campana08,campana13} is a cluster-detection algorithm useful 
for searching spatial concentrations in a field of points. 
As stated in the Introduction, we already applied this method to the $\gamma$-ray sky and 
detailed descriptions of the MST and of the selection criteria were presented elsewhere 
(e.g. in Paper I), therefore, we provide here only a brief summary of this method.

Consider a two-dimensional set of $n$ points or \emph{nodes}: several sets
$\{\lambda_i\}$ of weighted \emph{edges} connecting them can be defined. 
For a set of points in a Cartesian frame, the edges are the segments joining the nodes and 
the weights are their Euclidean lengths, while for a region on the celestial sphere the edge 
weights are the angular distances between pairs of photons. 
The MST is defined as the (unique) \emph{tree}, i.e. a graph connecting all the nodes without 
closed loops, that has the minimum total weight, defined as $\min [\Sigma_i \lambda_i]$. 

Once the MST is computed, by means e.g. of the Prim's algorithm \citep{prim57}, 
a set of subtrees corresponding to clusters of photons is extracted by means of a twofold 
selection, consisting firstly of a \emph{separation}, removing all the edges having a length 
$\lambda > \Lambda_\mathrm{cut}$. 
The separation value can be defined in units of the mean edge length 
$\Lambda_m = (\Sigma_i \lambda_i)/n$. 
This will result in a set of disconnected sub-trees; then, an \emph{elimination} will remove 
all the sub-trees having a number of nodes $N \leq N_\mathrm{cut}$, leaving only the clusters 
having a size over a fixed threshold. 
The remaining set of sub-trees provides a first list of candidate clusters and a \emph{secondary} 
selection is applied to extract the most robust candidates as $\gamma$-ray sources.
A suitable parameter for this selection \citep{campana13} is the \emph{clustering parameter} $g_k$ 
defined as the ratio between $\Lambda_m$ and $\lambda_{m,k}$, the mean length of the $k$-th 
cluster edges. 
Another very useful parameter for assessing the significance of the surviving clusters is the 
\emph{cluster magnitude}:
\begin{equation}
M_k = N_k g_k  
\end{equation}
where $N_k$ is the number of nodes in the cluster $k$ and the $g_k$ its clustering parameter. 
The probability to obtain a given magnitude value combines that of selecting a cluster with
$N_k$ nodes together with its ``clumpiness'', compared to the mean separation in the field.  
It was found that $\sqrt{M}$ is a good estimator of statistical significance of MST clusters 
\citep{campana13}.
In particular, a lower threshold value of $M$ around 15--20 would reject the large majority 
of spurious (low significance) clusters. 

The cluster centroid coordinates are obtained by means of a weighted mean of the photons' 
coordinates \citep{campana13}.
If the cluster can be associated with a genuine pointlike $\gamma$-ray source, the radius of the 
circle centred at the centroid and containing the 50\% of photons in the cluster, the 
\emph{median radius} $R_m$,  should be smaller than or comparable to the 68\% containment radius 
of instrumental Point Spread Function \citep[PSF, see][]{ackermann13b}.
Another useful parameter is the \emph{maximum radius} $R_\mathrm{max}$, defined as the angular 
distance between the centroid and the farthest photon, giving informations on 
the overall extension of the cluster. 
Table~\ref{t:params} reports a summary of the main parameters in the MST source detection algorithm.

\begin{table}[htbp]
\caption{Summary of the main parameters in the MST source detection algorithm and their meaning.}\label{t:params}
\centering
\begin{tabular}{cp{6cm}}
\hline
Parameter & Meaning \\
\hline
$\Lambda_m$				& 	Mean edge length in the Minimal Spanning Tree built on a given field.			\\
$\Lambda_\mathrm{cut}$  &	Cut length for the primary selection (e.g. defined as a fraction of $\Lambda_m$).			\\
$N_\mathrm{cut}$ & Minimum number of nodes for the primary selection.\\
$N$ & Number of nodes for a cluster.\\
$g$ & Clustering parameter, the ratio between $\Lambda_m$ in the field and the mean edge length in the cluster.\\
$M$ & Magnitude of the cluster, defined as $M=Ng$.\\
$R_m$ & Median radius, the distance from the cluster centroid containing 50\% of the photons.\\
$R_\mathrm{max}$ & Maximum radius, the distance from the cluster centroid to its farthest photon.\\
\hline
\end{tabular}
\end{table}%

\section{The MST catalogue}\label{s:data_analysis}
\subsection{The \emph{Fermi}-LAT dataset}
The full \emph{Fermi}-LAT dataset collected above 10 GeV in the 9 years from 2008 
August 4 to 2017 August 4, and processed with the Pass 8 reconstruction algorithm 
and responses, was downloaded from the FSSC 
archive\footnote{\url{http://fermi.gsfc.nasa.gov/ssc/data/access/}}. 
The standard cuts on data quality and zenith angle (\textsc{source} class events, front and back-converting, up to a maximum
zenith angle of 90\degr) were applied. 

\begin{figure}[htbp]
\centering
\includegraphics[width=0.5\textwidth]{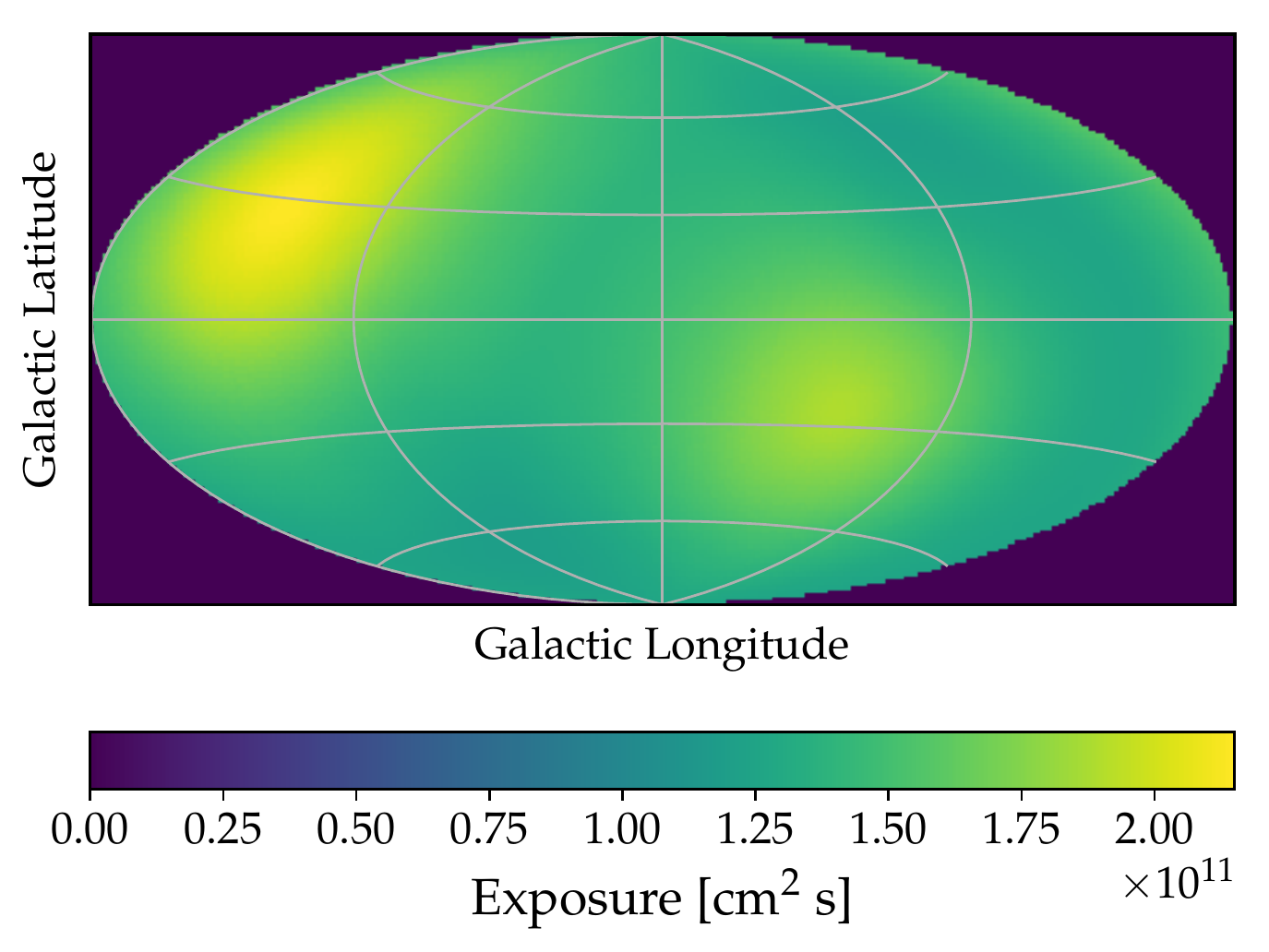}
\caption{Sky map of the Fermi-LAT exposure at 10 GeV in the first 9 years of observation
in the scanning mode.}
\label{f:expmap}
\end{figure}

The scanning mode of \emph{Fermi}-LAT resulted in a non-uniform exposure distribution 
over the sky with a maximum value near the North celestial pole and minimum exposure on 
the celestial equator, as apparent in the map reported in Figure~\ref{f:expmap}.

The detection of photon clusters, performed by means of MST, depends upon the mean 
spatial density of background events, that largely increases moving from the Galactic 
poles to the equator.
For this reason we removed from the data the region with $|b| < 19^\circ$, where
the clusters found (in particular those with a small number of photons) are poorly
stable, i.e. their number of photons changes drastically even for small variations 
of the $\Lambda_\mathrm{cut}$ parameter.
The final dataset contains 291,979 photons, of which 152,906 and 139,073 are in 
the North and South Galactic regions, respectively.
This difference is mainly due to the non uniform spatial distribution of the
exposure.

\begin{figure*}[htbp]
\centering
\includegraphics[width=\textwidth]{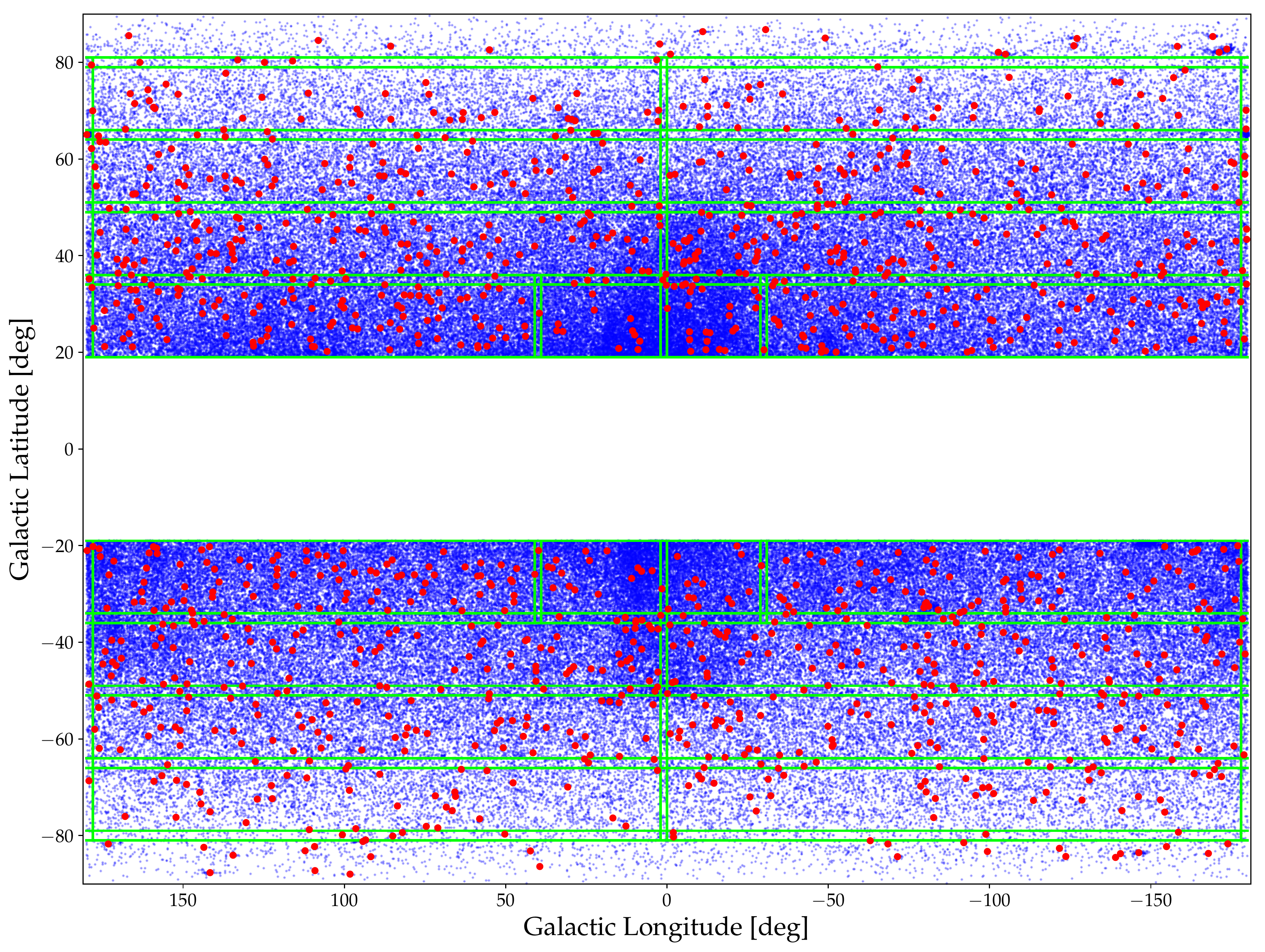}
\caption{Sky map of the photons in the 10--300 GeV energy range considered in our analysis.
Green lines mark the boundaries of the regions used for the MST cluster search.
Red dots are the clusters given in the final 9Y-MST catalogue.
}
\label{f:photonmap}
\end{figure*}

\begin{table*}[ht]
\centering
\caption{Sky regions used in the MST search for photon clusters.
Note the overlap on each side of the regions. 
The total number of photons and the resulting separation lengths for the North and 
South Galactic regions are given; the well evident N-S asymmetry is mainly due to the 
different exposures. 
In the last two columns are reported the number of clusters found in each region with
the primary selection and after the application of the secondary and superselection, as 
described in the text.
The Galactic longitude intervals are: A:~$0\degr < l < 182\degr$; B:~$178\degr < l < 362\degr$; C:~$0\degr < l < 41\degr$; 
D:~$39\degr < l < 182\degr$; E:~$178\degr < l < 331\degr$; F:~$329\degr < l < 362\degr$.}\label{t:regions}
\begin{tabular}{cccccccccc}
\hline
Region				&  Solid Angle    &     \multicolumn{2}{c}{Photon number}   & \multicolumn{2}{c}{$\Lambda_\mathrm{cut}$} & \multicolumn{2}{c}{$N_\mathrm{ps}$}  & \multicolumn{2}{c}{$N_\mathrm{ss}$}  \\
                             &     sr          &      N         &      S     &    N      & S  &   N & S   &   N & S  \\
\hline
$|b| > 79\degr$              &    0.1154       &  4639 & 3162   &  7\farcm4  & 13\farcm3  &  58 & 127 &  20 & 28 \\
\hline
$64\degr < |b| < 81\degr$ A  &    0.2824       & 10521 & 6428   &  6\farcm8  & 10\farcm1  & 102 & 138 &  45 & 42 \\  
$64\degr < |b| < 81\degr$ B  &    0.2955       &  9498 & 6014   &  7\farcm5  & 10\farcm5  & 105 & 154 &  42 & 44 \\ 
\hline
$49\degr < |b| < 66\degr$ A  &    0.5045       & 17267 & 14297   &  7\farcm4 &   8\farcm7 & 197 & 278 &  71 & 75 \\  
$49\degr < |b| < 66\degr$ B  &    0.5101       & 16361 & 12570   &  7\farcm9 &   9\farcm4 & 228 & 266 &  65 & 81 \\
\hline
$34\degr < |b| < 51\degr$ A  &    0.6923       & 24822 & 24121   &  7\farcm6 &   8\farcm0 & 425 & 512 & 110 & 93 \\  
$34\degr < |b| < 51\degr$ B  &    0.6999       & 23126 & 21469   &  8\farcm2 &   8\farcm4 & 564 & 435 & 107 & 86 \\
\hline
$19\degr < |b| < 36\degr$ C  &    0.1876       & 11769 & 10209   &  6\farcm1 &   6\farcm5 & 298 & 252 &  17 & 20 \\
$19\degr < |b| < 36\degr$ D  &    0.6544       & 25942 & 24282   &  7\farcm4 &   7\farcm8 & 466 & 510 &  94 & 89 \\
$19\degr < |b| < 36\degr$ E  &    0.7002       & 22555 & 25512   &  8\farcm4 &   7\farcm8 & 494 & 539 &  96 & 78 \\
$19\degr < |b| < 36\degr$ F  &    0.1510       &  9368 & 7060    &  5\farcm9 &   6\farcm8 & 208 & 147 &  24 & 15 \\
\hline
\end{tabular}
\end{table*}

\subsection{The primary selection}

To take into account the differences in the photon density, we divided the full non-Galactic 
sky into three strips 17$^\circ$ wide in Galactic latitude for each hemisphere, allowing for a 
2\degr\ overlap between the adjacent regions. 
The boundaries of these regions are illustrated in the sky map in Figure~\ref{f:photonmap}.
The strip nearest to the Galaxy was subdivided in four longitude sectors, while the other two 
strips were split in two sectors. 
The north and south Galactic polar regions ($|b| > 79^\circ$) were analyzed separately.

For each field the MST was computed on the unbinned photon maps applying a primary selection using 
$\Lambda_\mathrm{cut} = 0.7\,\Lambda_m$ and $N_\mathrm{cut} = 3$.
These values have been shown \citep{paperI,paperIV} to be rather optimal for the further 
selections.
The resulting cluster samples were merged and a first preliminary list of 6503 clusters 
was obtained.
The selected regions, their solid angles, photon numbers and the used absolute $\Lambda_\mathrm{cut}$ 
are summarized in Table~\ref{t:regions}; here it is also reported the number of clusters 
$N_\mathrm{ps}$ found in each region with the primary selection only.
Note that the separation length, when measured in arcmin, remains uniform in many regions
with the exception of the southern highest latitude belt and polar region, where it is
longer than 10\arcmin\ since these fields are more sparse than the northern counterpart, and therefore also larger than the PSF radius.
As a consequence of this fact, in these three regions there is a number of clusters 
much higher than in the corresponding northern regions, despite the lower photon numbers.

\subsection{The secondary selections}\label{s:secondarysel}
We applied firstly an uniform secondary selection over all the cluster dataset. 
A threshold value of $M>15$ has been assumed, while for the 4-photon clusters only
($N=4$) a more loose threshold of $M>12$ was applied to take into account all those 
having $g > 3$.
Moreover, multiple coincident clusters in the overlapping regions were eliminated 
taking only the ones with the highest $M$.
We obtained thus a list of 2153 candidate sources.

We then applied a further ``superselection'' using different criteria for different 
sky regions. 
This has been proved to be an efficient strategy to take into account the backgrond 
non-uniformities \citep[especially in the peri-Galactic regions and in the so-called 
Fermi Bubbles,][]{su10} and to increase the chance to select clusters corresponding 
to genuine candidate $\gamma$-ray sources.
Note that with the superselection we further restrict the analysis to Galactic latitudes 
$|b|>20\degr$.

The following four regions have been defined, each characterized by specific threshold 
on $M$ and $g$:
\begin{enumerate}
	\item  High latitudes and poles, $|b| > 50^\circ$: $M>18$ or $g >3.5$;
	\item  Middle latitudes, $30^\circ <|b|< 50^\circ$: $M>20$ or $g>4.0$;
	\item  External peri-Galactic belt, $20^\circ <|b|< 30^\circ$ with $0^\circ<L<330^\circ$: $M>22$ or $g>4.2$;
	\item  Central  peri-Galactic belt, $20^\circ <|b|< 30^\circ$ with $-30^\circ<L<30^\circ$: $M>24$ or $g>5.0$;
\end{enumerate}

Note that the ``or'' above is to be intended in Boolean sense, i.e. for the Galactic poles 
we select the clusters that have either $M>18$ or $g >3.5$.

\begin{figure}[htbp]
\centering
\includegraphics[width=0.5\textwidth]{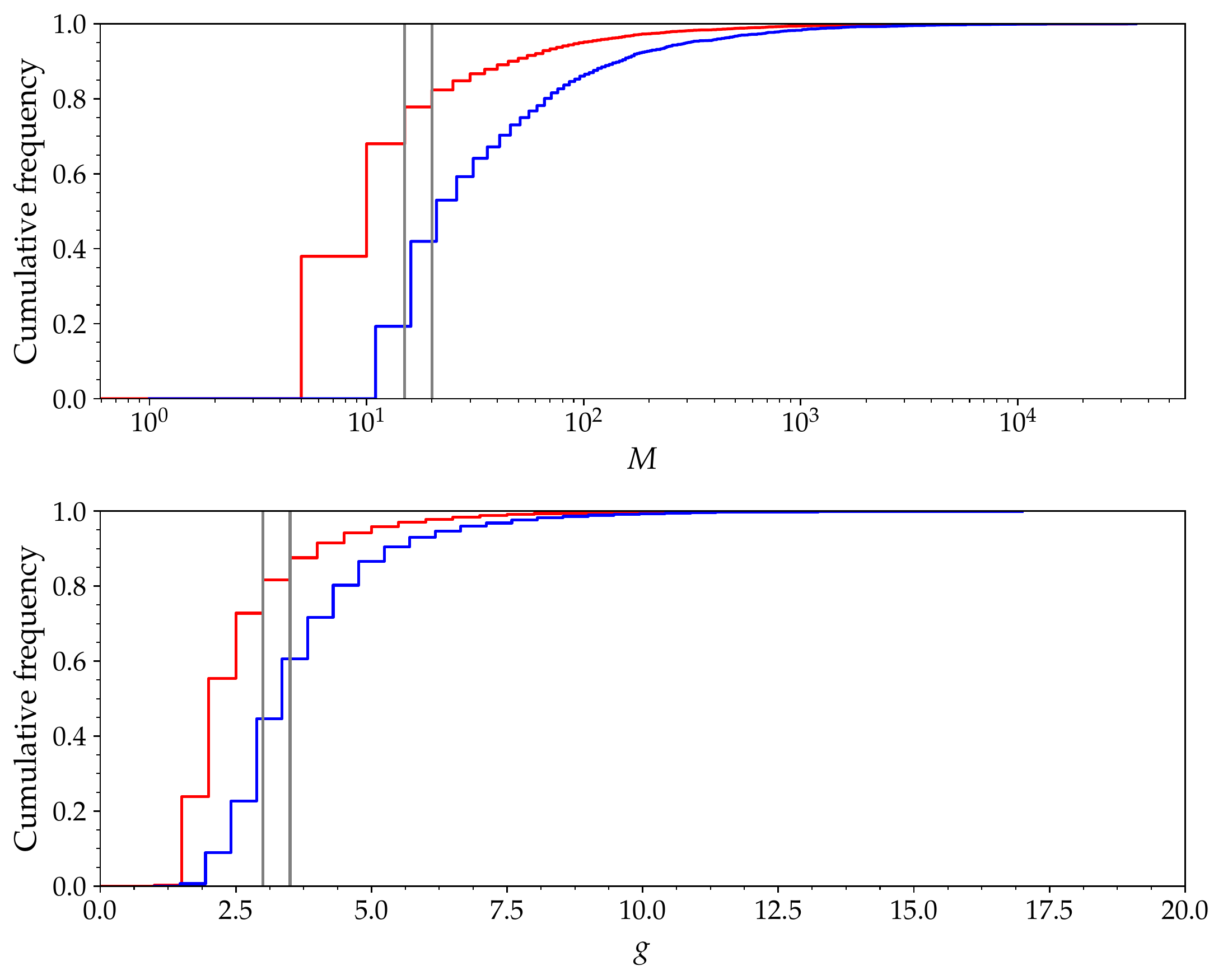}
\caption{\emph{Upper panel}: Cumulative distributions of the magnitude $M$ of the 
clusters after the primary (red) and secondary (blue) selections.
Vertical bars correspond to the cuts $M > 15.0$ and $M > 20.0$.
\emph{Lower panel}: Cumulative distributions of the clustering parameter $g$ of the 
same cluster sets.
Vertical bar corresponds to the selection thresholds  $g > 3.0$ and $g > 3.5$.
}
\label{f:gMcumdist}
\end{figure}

The choice of these superselection parameters was based on the experience from previous analyses, aiming at  
selecting a population having a large majority of significant structures, and will be further justified in the following Section~\ref{s:spurious}.
The cumulative distributions of the cluster parameters $M$ and $g$ are shown in the two panels
of Figure \ref{f:gMcumdist}, after the primary and the secondary selections.
In the figure some representative cut levels are marked, showing that the percentage of surviving clusters after 
both selections. 

The validity of our selection criteria will be further confirmed by the percentage of clusters with 
associations to known sources or other interesting objects, that resulted higher than 95\%, 
as it will be discussed in the following Sections.

After applying these criteria the number of candidate sources is reduced to 1342.
This number, about 20\% of the original primary selection clusters, is therefore comparable (i.e. about the same order of magnitude) to the amount of sources already detected in the same sky region, e.g. in the 3FHL or 3FGL catalogues.
The distribution of these clusters into the regions used in our analysis is also
given in Table~\ref{t:regions} as $N_\mathrm{ss}$, not including double and multiple
detections in the overlapping areas. 
It is interesting to note how the large excesses of clusters after the primary
selection in the regions with a low photon number is now largely reduced.
The ``superselection'' is therefore able to clean up the majority of low
significance strucures, up to about 90\% in some regions.
Of course, one cannot exclude that a fraction of clusters associated with genuine
high energy sources is also eliminated, but such a choice is a necessary trade-off
for obtaining a good sample of significant clusters.

\begin{figure*}[htbp]
\centering
\includegraphics[width=\textwidth]{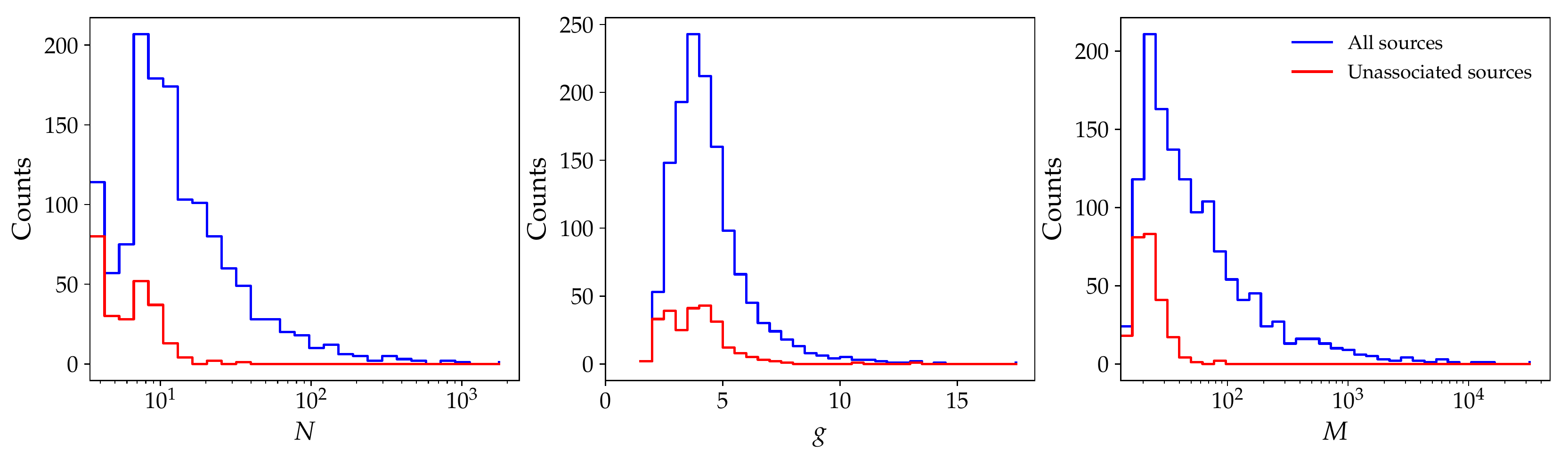}
\caption{Distribution of the $N$, $g$ and $M$ parameters for all the 9Y-MST sources 
(blue histograms) and for the unassociated sources only (red histograms, see main text for details).}
\label{f:NgMdist}
\end{figure*}

\begin{figure}[htbp]
\centering
\includegraphics[width=0.5\textwidth]{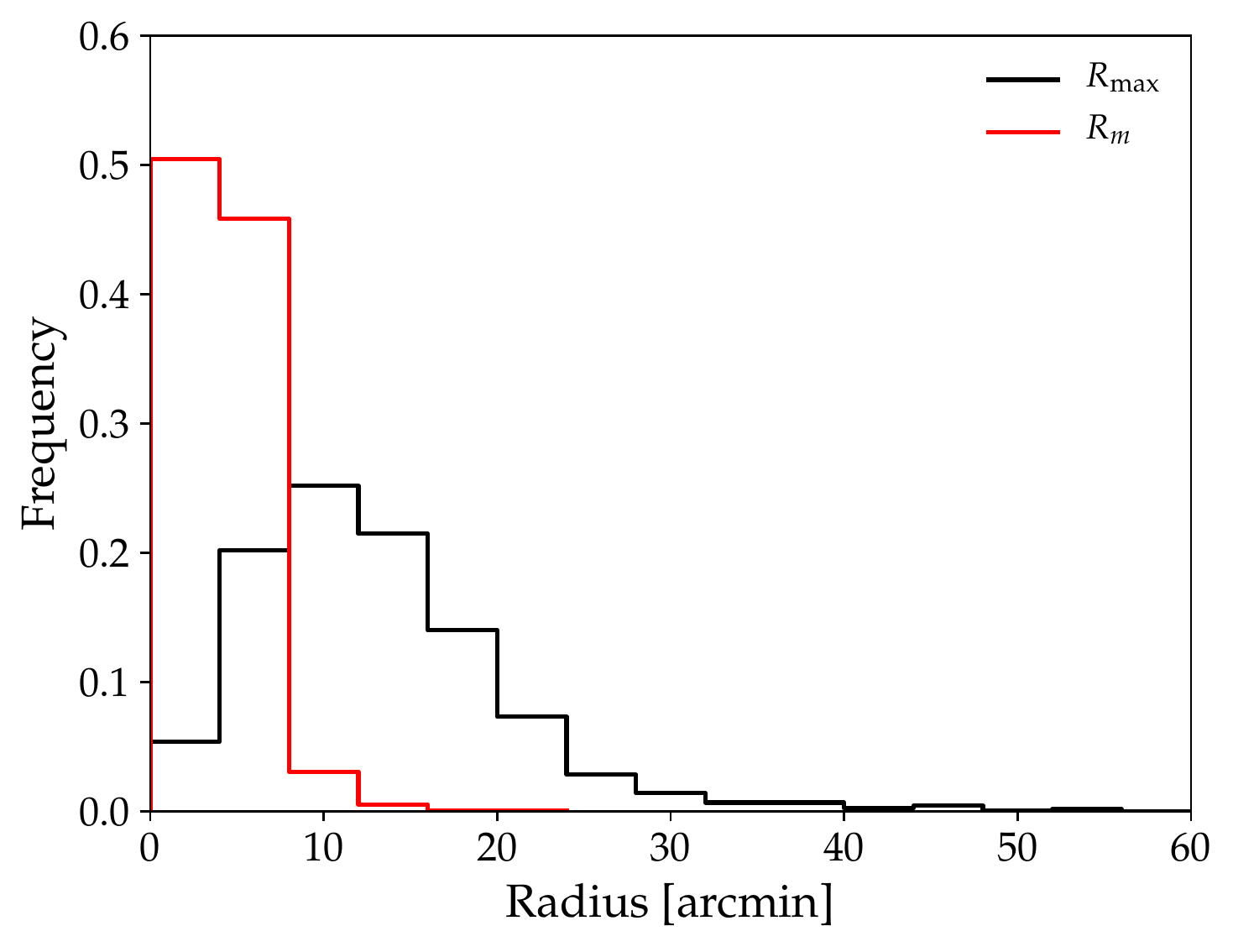}
\caption{Distributions of $R_\mathrm{max}$ (black histogram) and $R_m$ (red histogram) 
for all the 9Y-MST sources.
Note that how about all clusters have $R_m$ values smaller than 8\arcmin, while the other
distribution is peaked around 10\arcmin.
} 
\label{f:RmRmaxHis}
\end{figure}

These clusters are all included in the 9Y-MST catalogue FITS file available as 
additional electronic material\footnote{The catalogue is also available at the URL:\\ 
\url{http://www.iasfbo.inaf.it/~campana/9Y-MST_catalogue_latest.fits}}. 
The FITS contains cluster informations (coordinates, $N$, $g$, $M$, $R_m$ and 
$R_\mathrm{max}$) and the associated counterparts discussed in the following Sections. 
The sky map of the clusters is shown in Figure~\ref{f:photonmap}. 
The distributions of the $N$, $g$ and $M$ parameters for all the 9Y-MST sources are
shown in Figure~\ref{f:NgMdist}, while those of $R_m$ and $R_\mathrm{max}$ are given 
in Figure~\ref{f:RmRmaxHis}.

Note that there are a few very rich clusters ($N > 100$) and, in particular, only two
have more than 1000 photons and are associated with two well studied BL Lac objects:  
PG~1553+113 (5BZB~J1555+1111) with 1010 photons, and Mrk~421 (5BZB~J1104+3812) with
1997 photons.
The $R_m$ and $R_\mathrm{max}$ distributions are remarkably different, since the latter
is peaked around 10\arcmin\ and extends up to 55\arcmin, while more than 96\% of clusters
have $R_m < 8\arcmin$.

\begin{figure}[htbp]
\centering
\includegraphics[width=0.5\textwidth]{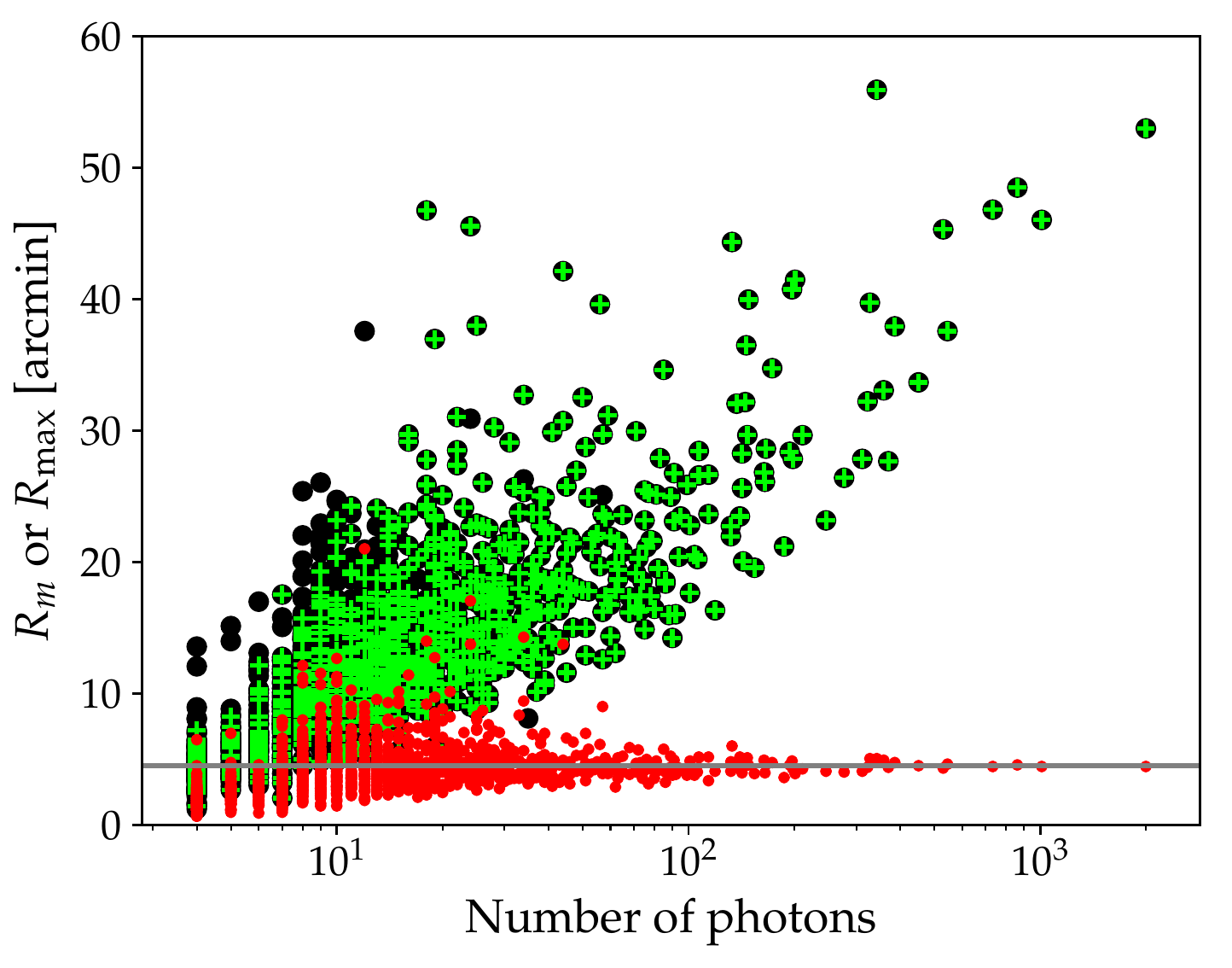} 
\caption{Scatter plots of the maximum radius $R_\mathrm{max}$ (black points)
and of the mean radius $R_M$ (red points) versus the number of photons in the clusters 
for all the 9Y-MST sources. 
Note that for high photon number the mean radius is very close the constant value 
of 4\farcm5 (horizontal line).
Green crosses overlaid on the black points mark the 3FGL counterparts to the 9Y-MST clusters.}
\label{f:NvsRmax}
\end{figure}

The scatter plots of the radii ($R_m$ and $R_\mathrm{max}$) versus the cluster photon number are shown in
Figure~\ref{f:NvsRmax}. Their difference is clear:
in particular, for the richest clusters $R_m$ is very close to 4\farcm5,
while the maximum radius is an order of magnitude larger.
This finding indicates that in the surroundings of large clusters there is a tendency
to find angular distances quite shorter than $\Lambda_m$ in the entire region,
with the aggregation of extended haloes.
An interesting case is that of clusters with a rather low number of photons but
a large $R_m$: they could be either extended structures or clusters having
a denser subcluster associated with nearby background photons.
In general these clusters are also characterized by a low clustering factor, and their properties will
be discussed in Section~\ref{s:lowgclusters}. 
Another item to be taken into account is the possibility to find small spurious clusters in 
the close surroundings of a very rich one: for this to happen, is enough to cut only one edge connecting 
the two structures.
The usual terminology is to call these \emph{satellite} clusters of the main one, 
see \citet{campana13}.
This matter will be further discussed in Section~\ref{s:satellites}.

\subsection{The occurrence of spurious clusters}\label{s:spurious}
The above discussed superselection rules were chosen and applied to reduce the possibility of including in the final
sample clusters originated by local fluctuations of the background, i.e. \emph{spurious} clusters
not corresponding to a real source.
Evaluating the expected fraction of spurious structures in the catalog is a complex task, 
but an estimate is useful for understanding the quality of the selection methods.
\cite{campana13}, on the basis of simulations with random fields, found that selections
with thresholds on $M$ ranging from 15 to 20 produce a low fraction ($\lesssim$10\%) of these spurious
clusters.
However, to simply use a high $M$ selection threshold does not give particularly interesting results,
since the most likely outcome is a sample containing bright sources that can be found with any other method.
Our main goal is to select clusters that can be associated with new sources: this implies that they can
also have a low photon number. For these reasons we applied also a selection on $g$.
Moreover, we have to consider the possibility to select structures with $M$ values above the
threshold but with low clustering as expected from extended or ``confused'' features.
\\
To evaluate the expected fraction of spurious clusters we performed numerical simulations
on a photon background similar to the observed one.
A simulated 9-years, $>$10 GeV \emph{Fermi}-LAT sky was generated using the \texttt{gtobssim} tool\footnote{\url{https://fermi.gsfc.nasa.gov/ssc/data/analysis/scitools/help/gtobssim.txt}}, including only the Galactic and isotropic diffuse backgrounds (\texttt{gll\_iem\_v06} and \texttt{iso\_P8R2\_SOURCE\_V6\_v06}) and assuming the same spacecraft attitude as the actual dataset.
The MST algorithm was applied using the same procedure and primary selection
parameters as for the real data: in particular, the absolute value of $\Lambda_\mathrm{cut}$, i.e. measured in degrees, was fixed 
to the one adopted in the selection on the true field.
Afterwards, the superselection criteria were applied also to the simulated data.
The resulting percentage of spurious clusters surviving to this procedure was $\sim$4\%. 
This imples that an estimate of the possible spurious cluster between the 1342 entries in the catalogue
gives a number of about 55 clusters, most of them with 4 or 5 photons.
However, slightly higher numbers of spurious clusters cannot be excluded, 
since the presence of many rich clusters in the true sky affects the MST construction in a way that it is difficult to reproduce in any simulation.
\\
Other spurious clusters can be present if there are significant photon density gradients, as observed in the regions near the Galactic belt, where it is possible to select low $g$ clusters close to the low $|b|$ boundaries where the photon density increases.
Not all these clusters may be spurious: some could be associated with real sources, 
considering that the number of Galactic $\gamma$-ray emitters can also increase at low latitudes.
In these cases, an analysis on small regions (typically having a size of $10\degr\times10\degr$) with decreasing 
$\Lambda_\mathrm{cut}$ values can help to verify whether these structures remain stable or they dissolve
into the background.
In the latter occurrence, this is a strong indication that the cluster could be spurious.
In the catalogue, a flag is reported in these $\sim$20 cases.

\section{Correspondence between 9Y-MST and the Fermi catalogues}

We searched for the correspondences between our 9Y-MST clusters with the two most recent main 
catalogues published by the Fermi collaboration: the 3FGL catalogue \citep{acero15}, 
including 3034 $\gamma$-ray sources detected in the energy range $ 0.1 < E < 300$ GeV, and 
the 3FHL catalogue \citep{ajello17},
including 1556 sources at energies higher than 10 GeV.
The former catalogue was based on the first 4 years of LAT data, while the latter one considered 
photons detected in a time interval of 7 years.
The sources in these two catalogues are distributed over all the sky, and those 
in the same field considered for the MST catalogue with $|b| > 20^{\circ}$ are 1715 
and 986 for the 3FGL and 3FHL respectively.

\subsection{3FHL catalogue}
\begin{figure}[htbp]
\centering
\includegraphics[width=0.5\textwidth]{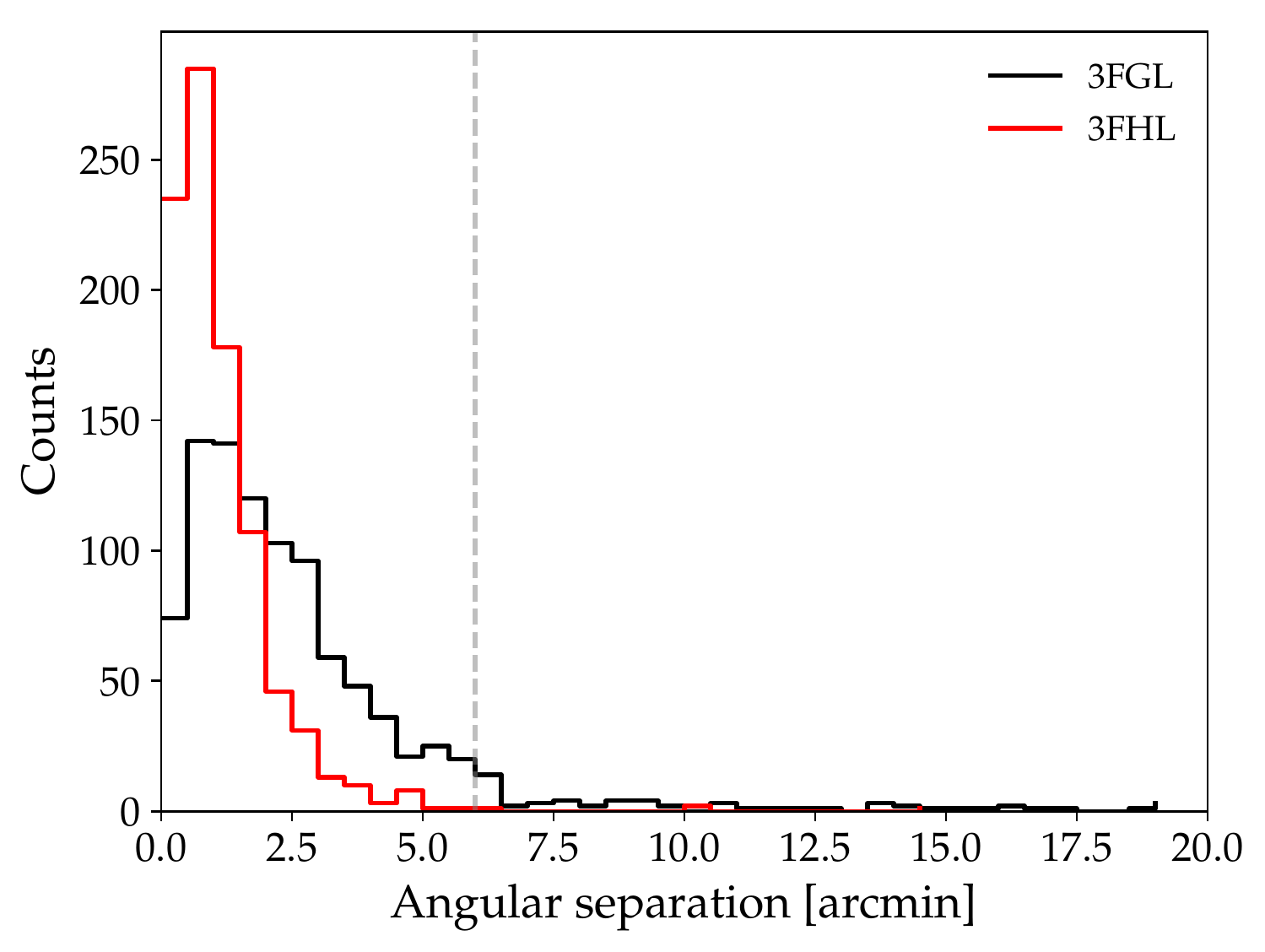}
\caption{Histograms of the angular distances $\delta$ between the clusters in the 9Y-MST 
catalogue and the sources in the 3FHL (red) and 3FGL (black) catalogues. 
Dashed vertical line marks the matching distance of 6\arcmin\ assumed for the 3FHL associations.}
\label{f:delta}
\end{figure}

We searched first the associations of 9Y-MST clusters with the 3FHL catalogue because of
their similar energy ranges, within a maximum separation of 20\arcmin\, and obtained 
923 
correspondences.
The distribution of the separations is shown by the red histogram in Figure~\ref{f:delta}.
About all the sources (916, 99.2\%) are found within a maximum distance $\delta$ of 6\arcmin.
We can conclude that this value can be adopted in other searches when at least one of the source positions 
is known with a much higher precision, as in the case of optical or radio catalogues.
In the search for associations between MST clusters and 3FHL sources the choice of a matching 
radius of 6\arcmin\ (0\fdg1) can be also based on the distribution of the semimajor axis of 
positional error ellipse at 95\% confidence of the 3FHL sources (``Conf\_95\_SemiMajor'' parameter 
in the 3FHL catalogue) which is smaller than this value for almost all  ($>$99.3\%) sources. 

For farther distances, at $\delta = 6\farcm5$ there is only one cluster associated with a 3FHL source: 
9Y-MST J0529$-$6904, with 34 photons and $R_\mathrm{max} = 26\farcm3$. 
This is an indication that it is clearly extended, being also located within the Large Magellanic Cloud.
No sources are found with $6\farcm5 < \delta < 10\arcmin$.
At larger angular separations, $10\arcmin < \delta < 20\arcmin$, we have 
6 possible matches. 
Five are multiple associations (i.e., there is another 3FHL source within 20\arcmin\ from the 
9Y-MST cluster, or \emph{vice versa}), and only one is a single association.
The latter is the
 cluster located at $RA = 269\fdg959$; $Dec=70\fdg612$ at an angular distance of
$14\farcm6$ from 3FHL J1757.7031 (corresponding to 3FGL J1756.9+7032) that has the BL Lac counterpart 
5BZB~J1757+7033; this angular distance appeared too large for a correct matching and we verified that
there is a cluster very close to the 3FHL source position with 7 photons and $M = 21.71$, just below
the ``superselection'' threshold for this region.
We, therefore, considered the cluster as not associated with the 3FHL source and included it in
the catalogue; we stress that this choice is strengthened by the fact that it is at $1\farcm1$
to the other BL Lac object 5BZB~J1759+7037.
The angular separation distribution thus indicates that the correct matching distance for the  
search of counterparts is within 6\arcmin\ and that the few sources found at higher separations 
are either random associations or peculiar occurrences.

It is clear that such a high fraction of associations cannot be due to chance correspondences
between the two catalogues.
However, it cannot be excluded that some associations might actually be spurious. 
We therefore evaluated how many correspondences are expected from a cross-matching of two catalogues having
similar large scale distributions.
We thus constructed a set of fake catalogues having the same number of 9Y-MST clusters
but coordinates shifted in RA and Dec by variable amounts ranging from 0\fdg5 to 3\fdg5.
The cross matching analysis with the true 3FHL catalogue within 6\arcmin\ gave a number of
correspondences ranging from 0 to 4, with a mean value equal to 1.5 and a standard deviation of 1.35.
Thus, it is reasonable to consider as genuine all 916 associations (92.9\% of the 3FHL catalogue 
sources).
For a further discussion on the matching radius, see also \cite{paperI}.

We then searched for the remaining 70 not associated 3FHL sources in the initial list of 2153
pre-superselection clusters, and 60 corresponding associations were found. 
These clusters were excluded from the 9Y-MST catalogue because they do not fulfill the
requested severe selection criteria. 
In particular, although nine of them had $M > 22.0$ or $g > 4.2$
they were however rejected being located in the region near the Galactic Bulge.

Therefore, there are only 10 3FHL sources not associated with MST clusters: all them 
are reported in the 3FHL with usually a rather low significance.
We can then conclude that $\sim$99\% of the 3FHL sources were found by our MST method, 
with a further $\sim$6\% rejected by the superselection criteria.

\begin{figure}[htbp]
\centering
\includegraphics[width=0.45\textwidth]{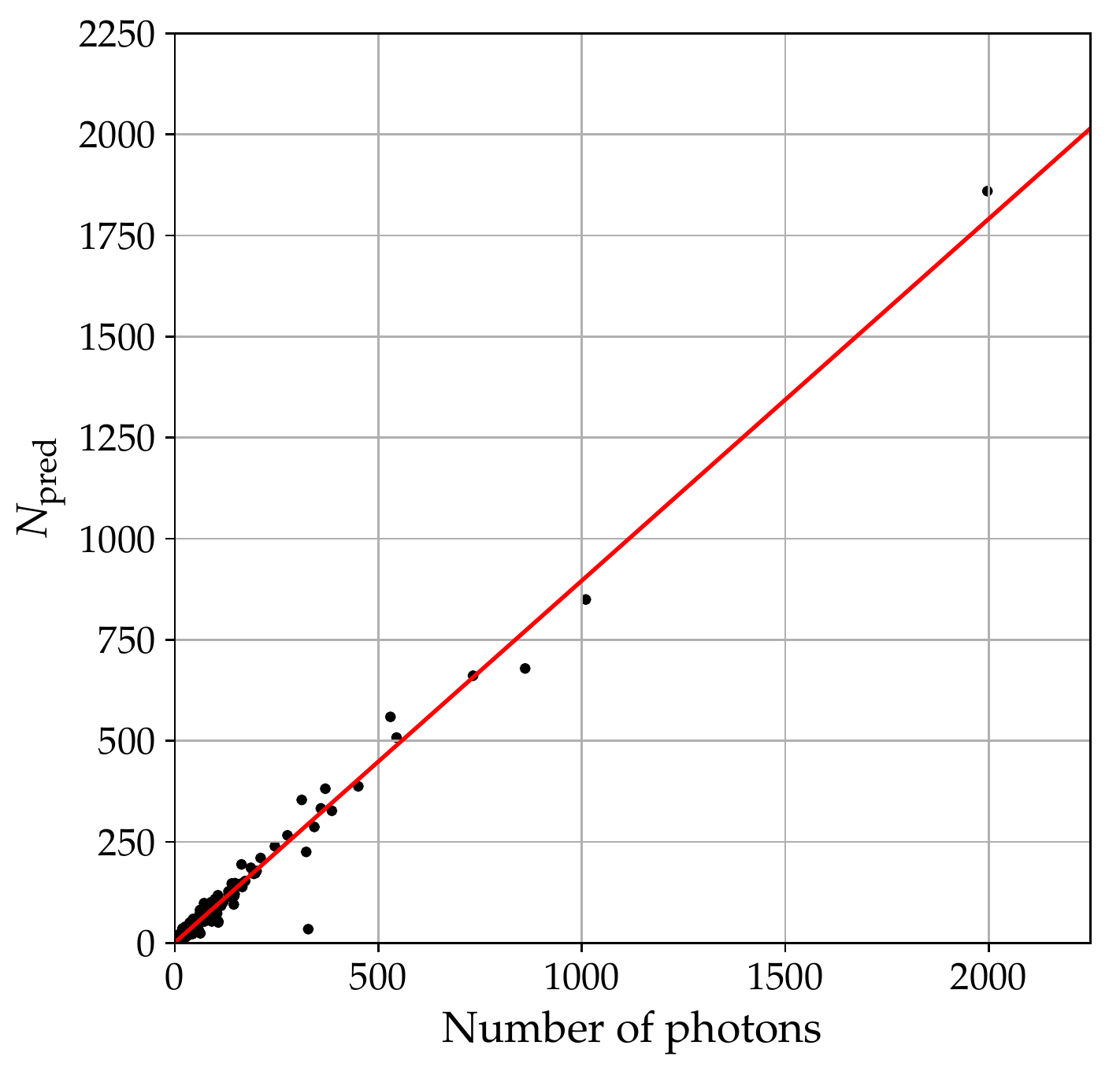}
\caption{Number of photons $N$ for the 9Y-MST clusters with a 3FHL counterpart versus 
the number of maximum likelihood predicted photons $N_\mathrm{pred}$.
The red line is the linear best fit.}
\label{f:N_vs_Npred}
\end{figure}

It is interesting to compare, for the matching MST-3FHL clusters, their number 
of photons $N$ with the number of predicted photons $N_\mathrm{pred}$ from the maximum 
likelihood analysis, reported in the 3FHL catalog. Figure~\ref{f:N_vs_Npred} shows this 
comparison: an excellent linear agreement with the two quantities is found,
with a slope $\sim$0.9 and an intercept compatible with 0 at the 2$\sigma$ level.
The only significant outlier is 3FHL~J2232.7$+$1143, for which a cluster with 328 
photons ($M\sim3120$) is found but the 3FHL catalog reports $N_\mathrm{pred} = 34.17$. 
This source (also known as CTA~102) had a significant outburst in late 2016, thus outside 
the 3FHL data range.

\subsection{3FGL catalogue}

The search for associations between MST clusters and the 1715 3FGL sources in the same sky region
resulted in 943 matchings within a maximum angular separation of 20\arcmin\ (55.2\%, Figure~\ref{f:delta}).
The fact that many sources of this catalogue are not associated with clusters is likely a 
consequence of the different energy ranges considered: the majority of the photons for the 3FGL 
sources have energies below 10 GeV, down to 100 MeV.
Moreover, the use of lower energy bands (characterized by a broader PSF) loosens the positional 
accuracy of the 3FGL sources, allowing for higher $\delta$ values. 
This is apparent from the $\delta$ distribution (black histogram in Figure~\ref{f:delta}), much 
broader with respect to the 3FHL sources.

In this case, we can consider $\delta = 15\arcmin$ as the threshold for a genuine association 
with the 3FGL catalogue. Note that about 96\% of the sources in the 3FGL catalogue have a positional 
error ellipse semimajor axis smaller than this value.
 
There are two MST sources that could be associated within 15\arcmin\ from the 3FGL J0849.3+0458 
source, 9Y-MST J0849+0456 at $\delta = 6\farcm2$ and 9Y-MST J0848+0507 at $\delta = 13\farcm2$. 
At a closer look to the photon distributions, it is clear that these are two distinct associations, 
with the 3FGL source associated with the closer MST cluster, while the farthest cluster has a 
very good association with a 3FHL source (3FHL J0848.7+0508).

There are only 8 sources with $15\arcmin < \delta < 18\arcmin$ 
It is possible that some of this group are spurious associations, and
therefore they will be examined in the next subsection.
Interestingly, 783 sources (i.e. about 83\% of the entire sample) are found within 
$\delta < 4\arcmin$, and 885 ($\sim$94\%) within $\delta < 6\arcmin$ confirming the accuracy 
of these positional estimates.

Thus, we consider as genuine 935 associations with the 3FGL catalogue (within 18\arcmin, 54.5\% 
of the sample).
Finally, note that only 792 of the 935 associated 3FGL sources are also in the 3FHL catalogue 
within a distance of 6\arcmin.
Thus, with the MST method, above 10 GeV, 143 3FGL sources not previously found in the 3FHL are detected. 

The total number of clusters in the 9Y-MST associated with either 3FHL or 3FGL is thus equal
to 1060.

\subsubsection{Possible association with 3FGL sources at large angular distances}

As mentioned above, 8 9Y-MST clusters have angular distances from possible 3FGL 
counterparts greater than 15\arcmin\ but smaller than 18\arcmin.

An effective way to investigate the robustness of these associations is to 
perform a MST analysis on smaller fields around the possible associations, 
usually a few degrees wide. For a likely genuine association, the centroid of 
the MST cluster shall become closer to the counterpart, and the cluster shall 
remain usually stable for different choices of $\Lambda_\mathrm{cut}$.

While one of these clusters (9Y-MST J0233$+$0656) has a much closer 3FHL 
counterpart, i.e. the more distant 3FGL association is very likely spurious, in three 
cases (9Y-MST J0009$-$1418, 9Y-MST J1202$+$3857 and 9Y-MST J1553$-$0305) 
the cluster remains at a wide or even increasing separation from the 3FGL 
counterpart for various choices of the surrounding field dimension and/or 
$\Lambda_\mathrm{cut}$.
We therefore discard these possible associations as spurious.

Only in two cases, 9Y-MST J1336$-$4043 associated with 3FGL J1335.2$-$4056 
and  9Y-MST J2127$-$3940 associated with 3FGL J2126.5$-$3926, the MST cluster 
separation from the 3FGL sources decreases (down to a few arcmin), thus we 
retain these associations. 

Finally, there are two 3FGL sources associated with 9Y-MST cluster pairs:
3FGL J1150.5$+$4155 and 
3FGL J0843.4$+$6713.
At a separation of 1\farcm48 from 3FGL J1150.5$+$4155, which has a well established
counterpart in the bright BL Lac object 5BZB J1150$+$4154, there is the quite rich cluster 
9Y-MST J1150$+$4154 with 86 photons and $R_\mathrm{max} = 18\farcm5$ while the much smaller
9Y-MST J1150$+$4209 with only 4 photons and a low $M = 17.27$ is located at 15\farcm28.
There are no interesting possible counterparts for the latter cluster, which can be reasonably
considered as a ``satellite'' of the former one.
The other case is that of 3FGL J0843.4$+$6713, without any reported counterpart,
which has two nearby clusters: 9Y-MST J0843$+$6713 ($N = 10$; $M = 27.87$) at the 
very small angular distance between the centroids of 8\arcsec, and 9Y-MST J0842$+$6656 
($N = 9$; $M = 32.03$) with $\delta = 17\farcm24$) and a close 3FHL association.
Such a large difference in the angular separations and the low probability ratio that 
the associated cluster is the more distant one ($\approx$$6 \times 10^{-5}$, estimated 
from the area ratio of the circles spanned by the two distances), 
indicates that the former one is the more likely associated with the 3FGL source.

\subsection{2FGL and 1FGL catalogues}
A further search for associations within 18\arcmin\ with the 2FGL catalogue 
\citep[][2 years of observations, 1024 sources in the sky region under investigation]{2FGL} 
and 1FGL catalogue \citep[][11 months of observations, 806 sources in 
the same sky region]{1FGL} revealed two further associations: 9Y-MST~J0723+5840 
associated with 1FGL~J0722.3+5837, and 9Y-MST~J0952+3931 associated with 
1FGL~J0952.2+392.

The total number of clusters in the 9Y-MST associated with either 3FHL or 3FGL/2FGL/1FGL 
is thus equal to 1064, implying that 79.3\% of the 9Y-MST sources have a counterpart 
in one of the \emph{Fermi}-LAT catalogues.

\section{Correspondence between 9Y-MST and other known $\gamma$-ray sources}\label{s:correspondence}

As mentioned in the Introduction, in a series of four papers we applied MST 
for searching in the LAT sky candidate sources likely associated with blazars at 
energies higher than 10 GeV.
In Paper I we searched in the 6.3 year LAT sky (Pass 7) clusters having possible
counterparts among the BL Lac objects reported in the 5th edition of the Roma-BZCAT
\citep{massaro14,massaro15} and found 19 clusters with $5 \le N \le 12$ and 
$15.2 \le M \le 36.47$. 
These sources were verified by means of the standard unbinned likelihood analysis 
that found 15 sources with ${TS} \ge 15$, with 7 sources having ${TS} \ge 25$.

In Papers II, III, and IV a similar analysis was performed on the 7 year LAT sky (Pass 8)
for searching clusters having association with the 1WHSP sample \citep{arsioli15},
with blazars of different type in the Roma-BZCAT, and with new blazar candidates with 
mid-IR selection, like WIBRaLS \citep{dabrusco14}, respectively.
These papers collectively reported 71 new clusters with $M > 20$, of which only 3 
resulted with likelihood test statistics ${TS} < 16$, and with 58 having ${TS} > 25$.
Furthermore, 67 of these 71 clusters were confirmed in the subsequent 3FHL catalogue.

In the 9Y-MST catalogue there are 73 clusters within an angular distance of 6\arcmin\ out of  
these 90 previously MST-found blazars (Papers I--IV); 3 more clusters are at distances of $\sim$6\farcm5, 
but their $R_\mathrm{max}$ is higher than 12\arcmin\ and one of them has also a $R_m$ as 
large as 8\farcm5: therefore, their associations cannot be excluded \emph{a priori}.
Of the remaining 14 non-associated clusters, 7 were reported in Paper I, 3 in Paper III and 4
in Paper IV.

We verified if some of these clusters were present in the lists produced with the primary
selection and found 12 with $M$ and $g$ values lower than the strict thresholds considered
for the superselection, although 8 of them had $M > 15$.
Only two clusters, both in Paper I with low ${TS}$ values (MST 0803+2440 and MST 1311+3951)
were not confirmed by the present analysis.
It is interesting that MST 1311+3951 is reported in the 3FHL (as 3FHL J1311.7+3954), 
together with another one of these 14 MST blazars (MST 1449+2746 in Paper III, 
corresponding to 3FHL J1449.5+2745).  

Therefore, 1073 of the 1342 9Y-MST sources were previously reported in either one of the 
\emph{Fermi}-LAT catalogues or in Papers I--IV.

Recently, \cite{arsioli17} reported the results of a search for $\gamma$-ray 
emission from blazars and candidates listed in the 2WSHP sample \citep{chang17} 
and detected a signal in the 0.3--500 GeV band from 150 objects (1BIGB sample) 
not reported in the 1FGL, 2FGL and 3FGL catalogues.
28 of the 1BIGB sources were in the previous lists found by MST analysis (Papers I--IV).
Of the 150 1BIGB sources, 51 of them are also associated with 9Y-MST clusters, 
within a distance of 6\arcmin.                              
Of these, 26 out of 28 correspond to Paper I--IV sources, while the other 
two are below the selection threshold.

After the subtraction of these sources we obtained a sample of 252 9Y-MST 
unassociated clusters.

Other three $\gamma$-ray sources were found to be associated with 9Y-MST clusters.
One is the famous GRB 130427A \citep{maselli14} clearly detected by Fermi-LAT 
\citep{ackermann14} and corresponding to a cluster of 16 photons ($M = 59.27$) 
located at 2\farcm35 from the GRB position.
The cluster 9Y-MST~2250$-$1255  is clearly associated with the flat spectrum 
radio source PKS 2247$-$131, without any known optical counterpart, that had a 
$\gamma$-ray flare in 2016 detected by Fermi-LAT \citep{buson16} and corresponds 
to a rich cluster with 25 photons ($M = 94.27$).
The third cluster is 9Y-MST J1544$-$0649 with 22 photons and $M = 96.6$ corresponding 
to the transient source Fermi J1544$-$0649, which brightened for two consecutive 
weeks beginning on 2017 May 15 \citep{ciprini17} and was later observed in other 
bands \citep{chornock17,kawase17}: 
it was associated with the galaxy 2MASX J15441967$-$0649156 that exibits some 
blazar properties.

Thus the final list of newly discovered clusters includes 249 objects and it is 
reported in Appendix, Table \ref{t:newsrc}.
Two of these sources are located in the LMC, a complex and interesting region. 
A discussion on the counterparts within this region is deferred 
to another paper \citep{campana18}.

The \emph{Fermi}-LAT collaboration has published on-line a preliminary list 
of $\gamma$-ray sources detected in 8 years of 
observation\footnote{\url{https://fermi.gsfc.nasa.gov/ssc/data/access/lat/fl8y/}}.
This FL8Y list will be superseded in the near future by a new official catalogue. 
It contains 5524 entries and 2931 of them are at Galactic latitudes $|b| > 20\degr$.
We verified how many 9Y-MST clusters have a positional correspondence 
with sources in this list and found 1177 matches within an angular distance 
of 8\arcmin, including all the correspondences with the 3FGL catalogue.
Moreover, in this residual sample we found four 3FHL, one 1BIGB sources and 
the above reported GRB, reducing thus the number of unassociated clusters to 159.
Thus, the percentage of clusters with a confirmed $\gamma$-ray counterpart 
can increase up to about 88\%.

\section{Particular cluster subsamples}

In the full 9Y-MST catalogue there are 449 clusters characterized either by a low number 
of photons or by a low clustering factor and consequently values of $M$ not high enough
to exclude that a fraction of them could not be associated with $\gamma$-ray sources,
but instead correspond to spurious structures.
In Section~\ref{s:spurious} we estimated that a percentage of 4\% or slightly higher of the total number of catalogue sources is indeed
expected to be not genuine.
Note also these subsamples contain about all unassociated clusters.
In the following Subsections we discuss the properties of these subsamples to achieve 
some specific indication whether they are or not originated by random localized fluctuations
of the photon density.

\subsection{Low $N$ clusters}
The number of clusters having less than 7 photons is 246, of which 114 have 4 photons,
57 with 5 photons and 75 with 6 photons.
A large fraction of these clusters is associated with known $\gamma$-ray sources and for several
others we found interesting possible counterparts to be confirmed by further investigations.
Thus 46 clusters with 6 photons are associated and 15 have candidate counterparts, 26
and 17 are those for the clusters with 5 photons and 33 and 30 are found among the 
4 photon clusters.
The low $N$ clusters for which we did not found any association or any candidate counterpart
are 82, that is 1/3 of the entire subsample.

\subsection{Low $g$ clusters}\label{s:lowgclusters}

There are 203 clusters in the 9Y-MST catalogue with a value of the clustering degree $g < 3.0$
and a minimum number of 7 photons; two are in the Large Magellanic Cloud and will not be
included in the following analysis, reducing their effective number to 201.
In principle, one can expect that this subsample might include a fraction of spurious
clusters, not directly related to pointlike sources but either to extended region of 
relatively high photon density or to blended pairs of close sources, or to single 
sources near some random concentrations of photons.
As discussed in Section~\ref{s:data_analysis}, the disentangling of these situations is not a simple task 
and a further MST analysis with a shorther $\Lambda_\mathrm{cut}$ in a rather small 
region surrounding the cluster under examination can provide useful information.

\begin{figure}[htbp]
\centering
\includegraphics[width=0.5\textwidth]{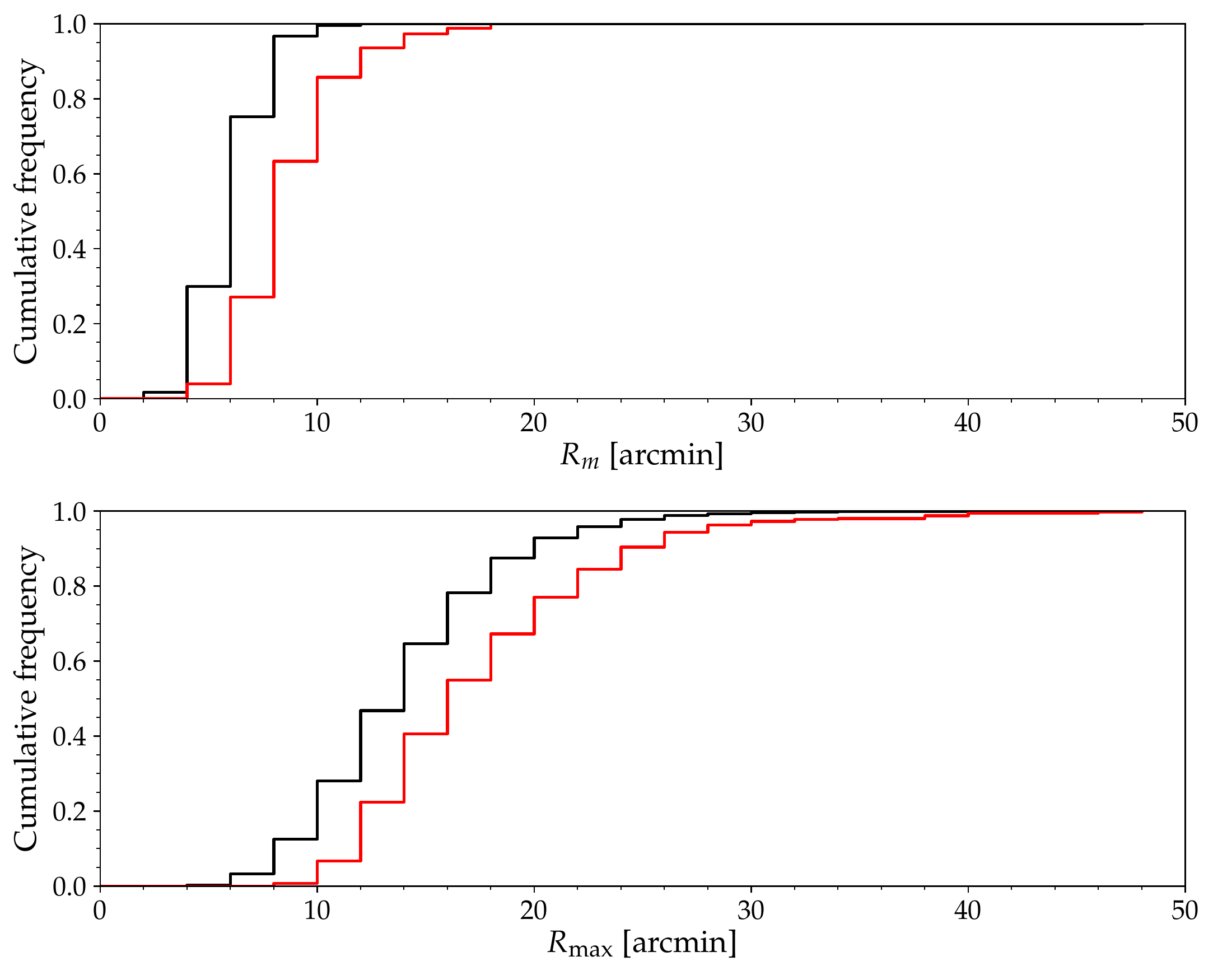}
\caption{Upper panel: Cumulative distributions of $R_m$ for the cluster samples with a
photon number in the interval [7, 25] and $g > 3$ (black line) and $g < 3$ (red line);
Lower panel: Cumulative distributions of $R_\mathrm{max}$ for the same two samples. }
\label{f:hisrrd}
\end{figure}

The values of $R_m$ and $R_\mathrm{max}$ of the low $g$ clusters are systematically higher than 
those of other clusters with a similar photon number but $g > 3$.
This is apparent from the cumulative distributions reported in Figure~\ref{f:hisrrd}.
Note, however, that more than 60\% of low $g$ clusters has $R_m < 8\arcmin$, a value
compatible with a point-like source; the large extension of clusters can be understood 
by the aggregation of a few photons of the local background due to the relatively long
local separation length.
The number of low $g$ clusters in the residual sample of unassociated entries, after 
the selection for correspondences with known catalogues and lists (see the previous Sections), 
is equal to 49, about 3.7\% of the entire 9Y-MST catalogue and 24\% of the subsample.
For 14 of these clusters, particularly those with a low $R_m$, we found that
they are located close to interesting radio sources and blazar candidates worthwhile of
further investigations.

Other clusters, when analyzed in a restricted field, dissolves into small unsignificant 
structures, and therefore can be related to an increase of the local background instead 
of a genuine source.
For example, in the case of the cluster 9Y-MST~J0009$-$3249 of 12 photons and the very large 
$R_\mathrm{max} = 37\farcm6$, at the Galactic latitude $b = -79\fdg365$ and at the angular 
distance of 4\farcm7 from a D$^3$PO structure \citep{selig15}, we found that it disappears
applying a $\Lambda_\mathrm{cut}$ value just below than used in the primary selection.

\section{Roma-BZCAT blazars in the 9Y-MST catalogue}\label{s:blazars}
  
Blazars of the two main types, i.e. BL Lac objects and Flat Spectrum Radio 
Quasars (FSRQ), are the richest class of high-energy $\gamma$-ray sources.
For this reason, it is interesting to investigate which clusters in the full 9Y-MST catalogue have 
a well established association with the known and candidate blazars in the 5th  
Edition of the Roma-BZCAT.
This catalogue contains 3561 radio detected AGNs divided into three main types: 
BL Lac objects, candidates and galaxy dominated blazars (BZB and BZG, 1428 sources), 
FSRQ (BZQ, 1909 sources), plus a rather small sample of sources of uncertain type 
(BZU, 224 sources).
The search for counterparts with all the entries in the catalogues gave 
726 associations within 6\arcmin\ and
738 associations within 8\arcmin, seven of which are double associations and are 
discussed in Section~\ref{s:multipleassbzcat}.
The mean angular separaton is $\langle \delta \rangle = 1\farcm65$, much lower
than the PSF radius at these energies, confirming the good estimates of the 
centroid coordinates given by our method.
Only 12 sources were found with $6\arcmin < \delta < 8\arcmin$, but for 10
of them their $R_\mathrm{max}$ were from $\sim$25\% to more than 3 times
larger than $\delta$, thus these associations cannot be excluded.

Two clusters (9Y-MST J2231$-$4422 and 9Y-MST J2358$-$4602) are 
located at $\delta$ values from the candidate counterparts comparable or 
larger than their $R_\mathrm{max}$, thus the possibility that they were spurious 
association due to random closeness must be taken into account.
Moreover, these two clusters have 3FGL sources at $\delta$ values of 5\farcm6
and 10\farcm5, respectively.

To verify the possibility of spurious associations we also searched 
for counterparts having an angular distance to the cluster centroid higher than
the corresponding $R_\mathrm{max}$.
In the sample of 738 associated blazars, only 3 did not satisfy this criterion:
one is 9Y-MST J2358$-$4602, reported above, and the other two are two clusters
of only 4 photons, very high clustering factors and $R_\mathrm{max}$ lower than 4\arcmin.
There is no statistical tool to establish if the associations of these two
clusters were spurious or not, but considering also that they were found  
positionally associated with FL8Y sources the most likely indication is for 
a genuine result.

The large majority of blazars related to $\gamma$-ray clusters are already in the 3FGL 
and 3FHL catalogues, while there are 34 associations with newly discovered 9Y-MST clusters 
with $\delta < 6\arcmin$, however 12 are already in the 1BIGB sample.
Thus, our new detections of known blazars above 10 GeV are 22.

Finally, our results confirm that the richest class of high energy $\gamma$-ray emitters
are BL Lac objects, with 548 sources, i.e. a percentage of about the 74\% of the total 
sample, while FSRQs are 155 ($\sim$21\%) and 35 ($\sim$4\%) belong to those with 
uncertain classification.
This finding is even more relevant in the subsample of Table~\ref{t:newsrc}, where the BL Lac objects
candidated as counterparts are 16, to which other 4 galaxy dominated must be added,
while the clusters with a possible association to FSRQ are only 3.

\subsection{Multiple associations with the Roma-BZCAT blazars}\label{s:multipleassbzcat}

As reported above, we found that seven 9Y-MST clusters can be associated with a pair of
5th Roma-BZCAT sources within the angular distance of 8\arcmin.
These rare cases can be due either to a single genuine association with a source in proximity
of another blazar or to a real confusion if both sources are $\gamma$-ray loud.

To discriminate between these two possibilities we performed a new MST analysis in 
small regions surrounding these clusters to verify if they are fragmented into more 
components when lower $\Lambda_\mathrm{cut}$ values are used.
For five of the seven clusters with a double blazar association, even performing a cluster 
search with $\Lambda_\mathrm{cut}$ values as low as $0.2\,\Lambda_{m}$, we obtained 
always single clusters with  $R_{m}$ and $R_\mathrm{max}$ smaller enough to exclude the more 
distant blazar as counterpart.
Only in the two cases of 9Y-MST J0009+0627 and 9Y-MST J0442$-$0019 the source confusion 
remains, because the possible counterparts are found quite close to the cluster centroids
even with the shortest $\Lambda_\mathrm{cut}$ values.
More precisely, the former cluster with $R_{m} = 3\farcm1$ and $R_\mathrm{max} = 10\farcm2$ has
two blazars 5BZQ J0009+0625 and 5BZB J0009+0628 at $\delta$ equal to 2\farcm1 and 2\farcm6,
respectively, and the latter ($R_{m} = 6\farcm1$, $R_\mathrm{max} = 20\farcm6$) has 
5BZB J0323$-$0111 and 5BZB J0323$-$0108 at 1\farcm5 and 5\farcm6, respectively.

\section{High-energy unassociated clusters}

As reported in Section~\ref{s:correspondence}, after having removed the clusters 
corresponding to the high-energy associations discussed in the preceding Sections, we 
obtained 249 clusters not corresponding to any source in the published Fermi 
catalogues or in the LMC (but including 89 new sources in the preliminary FL8Y list).

We also searched for positional correspondences between these clusters and objects in
catalogues of possible blazars selected on the basis of the occurrence of some features 
typical of this class of active galactic nuclei.
The catalogues used were CRATES \citep{healey07}, that reports 11,131 flat spectrum radio 
sources, WIBRaLS \citep[WISE Blazar-like Radio-loud Sources][]{dabrusco14}, listing 7855 
sources with mid-infrared colours typical of blazars, and 2WHSP \citep{chang17}, a multi-frequency 
catalogue of 1691 high energy and very high energy $\gamma$-ray blazars and blazar candidates.
Considering that in all the three catalogues the source coordinates are known with a high
accuracy, we considered angular distances lower than 6\arcmin\ for possible associations with 
$\gamma$-ray clusters, in agreement with the angular distance distribution given in 
Figure~\ref{f:delta}.
The 22 BZCAT objects already found in the previous search were excluded to avoid 
redundant associations (Section~\ref{s:blazars}).

The resulting associations were: 15 clusters with CRATES sources, 14 with WIBRaLS, and 15
with 2WHSP, but some are common to two or to all three samples.
The final number of clusters associated with entries of any of these three catalogues is 28
and they are indicated by a specific note in Table~\ref{t:newsrc}.

\subsection{Extended and satellite clusters}\label{s:satellites}

The nature of the remaining unassociated clusters can be somewhat complex and several effects
must be taken into account.
As discussed in Section~\ref{s:secondarysel}, there are several clusters having values of 
$R_m$ and $R_\mathrm{max}$ in excess with respect to those generally found for clusters 
associated with point-like sources.
In Table~\ref{t:newsrc}, large size clusters likely corresponding with extended structure are indicated by 
``ex''.
When investigated in rather small regions and with decreasing separation lengths,
several clusters have an inner compact structure with radial values in agreement with the typical
ones for clusters with a comparable photon number.
We indicated them by the note ``cc+ex''.
In the same analysis other clusters dissolve in non-significant structures and likely should be
considered spurious, and for this reason they have the annotation ``sp''.

Another interesting possibility is the occurrence of possible \emph{satellite} clusters, 
i.e. rather small clusters found at a relatively short distance from a much richer cluster.
In principle, the nodes (i.e. the photon arrival directions) might belong to a unique cluster, 
but they were partitioned in two because of the occurrence of only one edge slightly longer 
than the separation distance. 
A satellite cluster usually has a $g$ value and a photon number much lower that those of the 
nearby main cluster.
Other possibilities are the occurrence of an extended emission, or of a single weak cluster embedded
in a relatively high photon density region, and of a pair of close weak clusters.

We searched for satellites applying the simple criterion of sorting all cluster pairs
having an angular separation of their centroids lower than the sum of their $R_\mathrm{max}$.
Close weak clusters are expected to be resolved by a local MST analysis with a shorter 
$\Lambda_\mathrm{cut}$, while an extended emission will preferably fragment in a few 
low significance features.
 
\begin{figure}[htbp]
\centering
\includegraphics[width=0.5\textwidth]{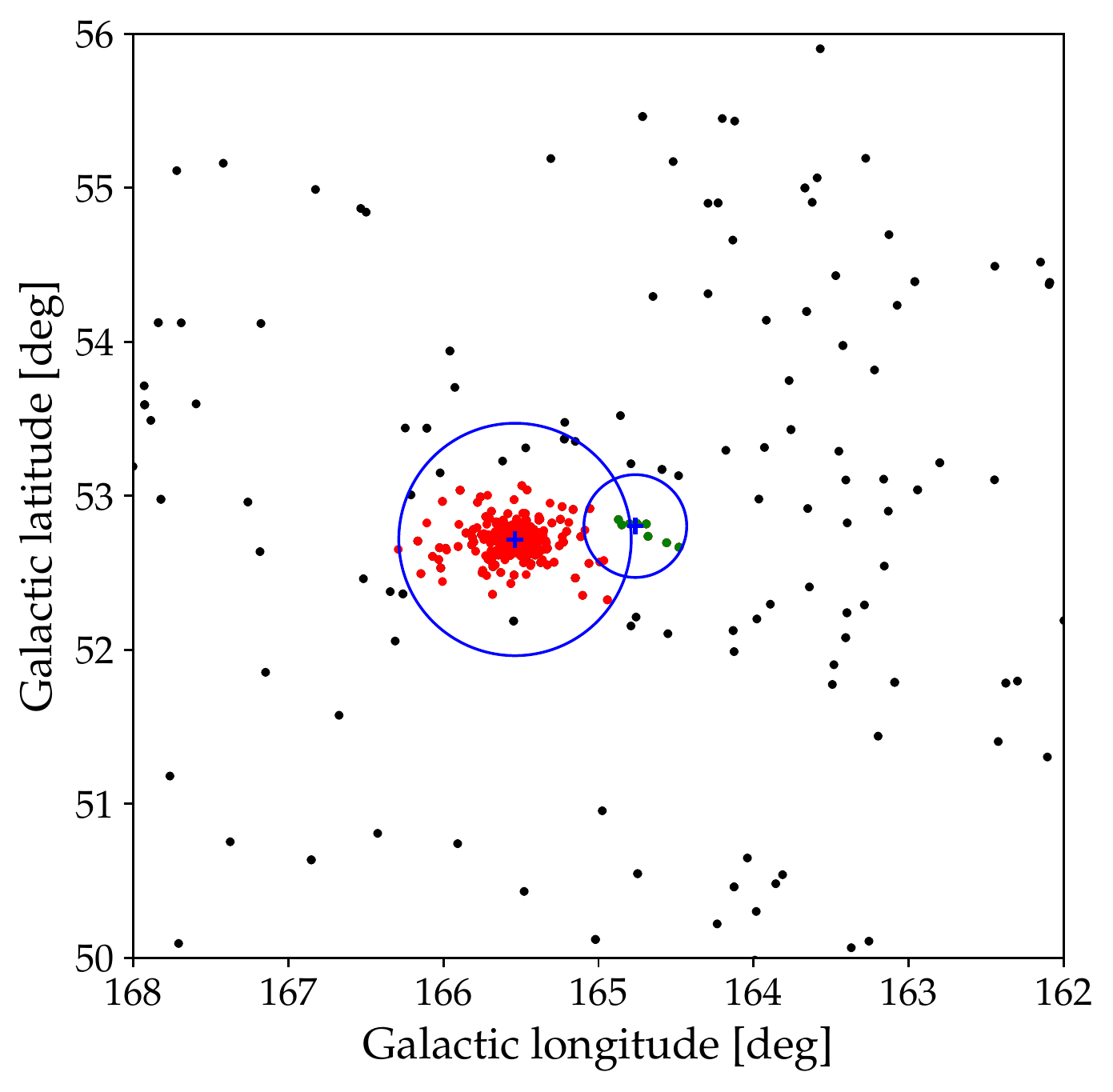}
\caption{Photon map at energies higher than 10 GeV in the sky region around the bright BL Lac
object 1ES 1011+496.
Photons in the rich cluster corresponding to the source are reported in red while the green
points are the 8 photons in the cluster 9Y-MST J1016+4949, likely a satellite of the former.
Crosses indicate the centroid positions and the circles' radii represents their $R_\mathrm{max}$.
}
\label{f:satel}
\end{figure}

The results of this search include two triplets, one of which (already mentioned above)
in the region of the Large Magellanic Cloud. 
The second triplet is in the enviroment of the rich cluster 9Y-MST J1512$-$0906 (343 photons) 
associated with a 3FGL/3FHL source and with the bright FSRQ PKS 1510$-$089.
The two other clusters are 9Y-MST J1508$-$0904, an unassociated low $g$ cluster (thus a 
reliable satellite candidate) and the much more distant 9Y-MST J1514$-$0948, 
close to a 3FHL source and to one of the previously detected MST blazars (Paper III).
We found 9 pairs of poor clusters close to much richer ones and 8 of them are unassociated.
An example is shown in Figure~\ref{f:satel}, in which the photons in a rather small 
sky region around the BL Lac object 1ES 1011$+$496 and the two MST-found clusters are plotted: 
the richer cluster (370 photons) is the one associated with the blazar, while a much smaller 
unassociated cluster is found within the $R_\mathrm{max}$ distance.

The rich cluster 9Y-MST J2139$-$4235 with 107 photons and $R_\mathrm{max} = 39\farcm5$, has the 
companion 9Y-MST J2138$-$4312 at 38\farcm2.
The latter one appears a significant structure ($N = 8$ and $M = 30.08$) instead of a  
satellite; moreover, it is at 4\farcm4 to a X-ray loud AGN with blazar mid-IR colours,
enforcing this conclusion.

There are also 6 pairs with both clusters well associated with already known $\gamma$-ray 
sources.
These cases must be considered as resolved with the $\Lambda_\mathrm{cut}$ employed.
The 4 remaining pairs have unassociated clusters with a rather low number of photons: their 
classification as satellites is therefore uncertain, because the occurrence of
extended regions or of low significance close features can not be excluded.

\section{Summary and discussion}\label{s:conclusions}

We analysed the first 9 years of \emph{Fermi}-LAT data, using the Pass 8 events 
at energies higher than 10 GeV by means of the MST algorithm for detecting photon clusters.
In the selection procedure we adopted severe threshold values to reduce the possibility 
of spurious detections due to local background fluctuations.
We limited our search to the Galactic regions with $|b| > 20\degr$ to avoid confusion 
problems in the Galactic belt due to its high photon density, that makes difficult the 
choice of the best separation length.

A new catalogue of 1342 clusters was obtained, the large majority of which has a close
spatial correspondence with $\gamma$-ray sources reported in the Fermi collaboration 
catalogues, but with new sources detected in different multiwavelength searches and 
with some others in the more recent preliminary FL8Y list.
With respect to the 3FHL catalogue, which has 986 sources in the regions with a Galactic 
latitude higher than 20\degr, the number of 9Y-MST clusters is increased by about 36\%. 
The sample of new detections contains 249 entries, and is reported in Table \ref{t:newsrc}.
Of these, 89 are also in the FL8Y list, thus there is a residual sample of 160 new 
$\gamma$-ray clusters.
The search in catalogues of objects dominated by non-thermal emission in different
electromagnetic bands gave 46 possible associations.
We also searched for new possible blazar candidates within a region centred at the cluster
centroid coordinates and having a radius of 6\arcmin.
The search was based on possible optical or IR counterparts of radio sources, when present,
or of quasars or candidates reported in large databases. 
Thus, 53 candidate objects were selected, whose blazar nature cannot be considered
as confirmed since optical spectra are generally not available and in some cases there
are no radio detections because their flux densities could be lower than the survey limits.
A more detailed analysis of these objects will be presented in a forthcoming paper.
Eventually, the number of clusters without any confirmed or possible counterpart is 62, of which 
19 are classified as possible satellites or spurious clusters, i.e. less than 5\% of the 
catalogue, a figure that confirms the efficiency of the method and of the selection 
criteria.

As known from other high energy catalogues, the large majority of clusters is associated
with blazars, and particularly with BL Lac objects (see Section~\ref{s:blazars}).
The continuos observation of the entire sky by \FLL over period of about a decade can
enrich our knowledge on the population properties of this class of AGNs.
It can be expected that the study of faint $\gamma$-ray sources can lead to the discovery
of semi-quiescent or rarely flaring blazars, i.e. galactic nuclei exhibiting only
occasionally bright high energy bursts and remaining for rather long times in states too
faint for allowing their inclusion in sample based on a few ``shot-like'' measurements.

\begin{acknowledgements}

We acknowledge use of archival Fermi data. We made large use of the online version of 
the Roma-BZCAT and of the scientific tools developed at the ASI Science Data Center (ASDC),
of the final release of 6dFGS archive,
of the Sloan Digital Sky Survey (SDSS) archive, of the NED database and other astronomical 
catalogues distributed in digital form (Vizier and Simbad) at Centre de Dates astronomiques de 
Strasbourg (CDS) at the Louis Pasteur University.
\end{acknowledgements}

\bibliographystyle{aa}
\bibliography{bibliography} 

\clearpage
\onecolumn

\begin{landscape}
\longtab{3}{
\begin{longtable}{lrrrrrrrrrl}
\caption{\label{t:newsrc}  
The sources in the 9Y-MST catalogue unassociated with any \emph{Fermi}-LAT catalogue. See main text for details. 
Column $N$ gives the number of photons in the MST cluster, $g$ its clustering degree and $M = Ng$ its magnitude.
Meaning of the annotation is the following: FL8Y:    cluster with a correspondent source in the FL8Y list within 6\arcmin; LMC:     cluster in the Large Magellanic Cloud; BZc:     possible blazar candidate; WB:      possible counterpart in the WIBRaLS sample; CR:      possible counterpart in the CRATES catalogue; 2W:      possible counterpart in the 2WSHP catalogue; 5BZB:    cluster with a correspondent BL Lac object in the 5BZCAT; 5BZQ:    cluster with a correspondent FSRQ in the 5BZCAT; 5BZG:    cluster with a correspondent galaxy dominated blazar in the 5BZCAT; AGN:     possible AGN counterpart ; sp:      possible spurious cluster due to background fluctuations; spp:     very likely spurious cluster due to background fluctuations; ps:      possible satellite cluster; ex:      extended cluster; c+ex:    extended cluster with interesting core structure; cc+ex:   extended cluster with compact core; ccc+ex:  extended cluster with significant compact core.}\\ \hline\hline
Name                & RA (J2000)       & Dec (J2000)     & $l$        & $b$        &  $N$   & $g$     & $M$        &  $R_m$    & $R_\mathrm{max}$   & Notes \\
	                & \degr       & \degr     & \degr        & \degr        &     &      &         &    \arcmin  &  \arcmin  &  \\
\hline
\endfirsthead
\caption{continued.}\\\hline\hline
Name                & RA (J2000)       & Dec (J2000)     & $l$        & $b$        &  $N$   & $g$     & $M$        &  $R_m$    & $R_\mathrm{max}$   & Notes \\
	                & \degr       & \degr     & \degr        & \degr        &     &      &         &  \arcmin    & \arcmin   &  \\
\hline
\endhead
\hline
\endfoot
9Y-MST J0000$-$0216 &   0.104  & $-$2.272  &  94.596  &  $-$62.303 &  4   & 4.398 &   17.594 &  2.8     & 5.9    &  BZc                                                                         \\
9Y-MST J0009$+$0855 &   2.364  &  8.923    &  105.723 &  $-$52.540 &  5   & 5.909 &   29.544 &  1.6     & 4.1    &                                                                              \\
9Y-MST J0009$-$1418 &   2.379  & $-$14.306 &  83.530  &  $-$73.878 &  4   & 3.540 &   14.160 &  3.1     & 7.2    &  FL8Y                                                                        \\
9Y-MST J0009$-$3249 &   2.482  & $-$32.828 &  358.036 &  $-$79.365 &  12  & 2.002 &   24.025 &  21.0     & 37.6  &  sp, ex                                                                      \\
9Y-MST J0012$-$0429 &   3.037  & $-$4.487  &  98.725  &  $-$65.495 &  9   & 2.057 &   18.516 &  10.7     & 18.4  &  sp, ex                                                                      \\
\end{longtable}
}
\end{landscape}
\end{document}